\documentclass[onecolumn,amsmath,amssymb]{qm2008_abs}
\usepackage{graphicx}
\usepackage{dcolumn}
\usepackage{bm}
\usepackage{qm2008_symb}
\usepackage{multirow}
\newcommand{\pt}{p_\mathrm{T}}

\begin{document}

\title{{\Large Open charm measurement with HFT at STAR }}

\bigskip
\bigskip
\author{\large Jan Kapit\'an (for the STAR Collaboration)}
\email{kapitan@rcf.rhic.bnl.gov}
\affiliation{Nuclear Physics Institute ASCR, Rez/Prague, Czech Republic}
\bigskip
\bigskip

\begin{abstract}
\leftskip1.0cm
\rightskip1.0cm

Thermalization is one of the key questions in understanding the matter created in Au+Au collisions at RHIC. Heavy-flavor quark collectivity could be used to indicate the degree of thermalization of the light-flavor quarks. 
Heavy quark energy loss could give important information on color charge density of the medium. Direct reconstruction of open charm hadrons is essential for these measurements.

The Heavy Flavor Tracker (HFT) is a proposed upgrade of the STAR experiment. Full GEANT simulation of tracking with HFT was performed, showing its excellent capability to reconstruct open charm hadrons in broad $\pt$ range at midrapidity. Estimated errors on measurement of $\mathrm{D}^{0}$ meson $v_{2}$ and $R_\mathrm{CP}$ are presented.

\end{abstract}
\maketitle

\section{Introduction}

Heavy-flavor quark (c,b) production is one of the key measurements for systematic study of the dense medium created in heavy-ion collisions at RHIC \cite{p5}. 
Produced by initial hard scattering, heavy quarks can be used to study early stages of the collision. They derive their mass from the Higgs field, and therefore are not modified by the surrounding QCD medium.

Previous studies have identified the development of partonic collectivity in heavy-ion collisions at RHIC \cite{p5}, but they have not yet demonstrated thermalization of the created matter. 
The study of heavy quark collectivity may allow us to address this issue. Measurement of elliptic flow ($v_{2}$) of open charm hadrons to low transverse momentum ($\pt$) is of particular interest.

Suppression of high $\pt$ hadron production at RHIC \cite{jaro1,jaro2,jaro3} is commonly thought to arise from partonic energy loss in dense matter due to induced gluon radiation \cite{jaro4}. Radiative energy loss of heavy quarks is expected to be suppressed (dead cone effect) \cite{jaro11}.

Currently most of heavy-flavor analyses at RHIC use semileptonic decay modes of heavy-flavor hadrons, measuring decay electrons. Lack of precise kinematical information about the parent hadron makes it difficult to study dynamics at low $\pt$~\cite{po1}.
Measurements of $R_\mathrm{AA}$ of heavy-flavor decay electrons at high $\pt$ \cite{jaro,phenix} indicate a significant energy loss of heavy quarks. However, knowledge of the relative contributions of charm and bottom decays to electron spectra is crucial to interpret these results.

For the precision study of heavy-flavor quark collectivity and energy loss, direct reconstruction of heavy-flavor hadrons is necessary. Given the large combinatorial backgrounds in heavy-ion collisions, topological reconstruction is needed to achieve reasonable signal significance.
The Heavy Flavor Tracker (HFT), a proposed upgrade \cite{po4} to the STAR experiment, will enable measurement of open charm hadrons by reconstructing their displaced decay vertices. 

\section{Heavy Flavor Tracker design}

The HFT detector consists of two subsystems: The PIXEL detector (2 layers) and Intermediate Silicon Tracker (IST, 1 layer). The midrapidity tracking system of STAR further includes existing Silicon Strip Detector (SSD) and the large Time Projection Chamber (TPC).
Compared to the previous version, the current design of the HFT has been optimised for lower mass and better hit resolving, however it has not been fully simulated yet. In the following, simulation results are presented for the previous design. The parameters of these are displayed in Table~\ref{tab:parameters}.

\begin{table}[htb]
\begin{center}
\begin{tabular}{|l||c|c||c|c|}\hline
& \multicolumn{2}{c||}{current design} & \multicolumn{2}{c|}{simulated design} \\ \hline
\multirow{2}{*}{layer}& \multirow{2}{*}{r (cm)} &Hit resolution & \multirow{2}{*}{r (cm)} & Hit resolution \\
& & ($r-\phi \times z$) ($\mu\mathrm{m} \times \mu\mathrm{m}$) & &($r-\phi \times z$) ($\mu\mathrm{m} \times \mu\mathrm{m}$) \\ \hline \hline
SSD & 23 & 30 $\times$ 699 & 23 & 30 $\times$ 699 \\ \hline
IST2-B & - & - & 17 & 17 $\times$ 12000 \\ \hline
IST2-A & - & - & 17 & 12000 $\times$ 17 \\ \hline
IST1   & 14 & 115 $\times$ 2900 & 12 & 17 $\times$ 6000 \\ \hline
PIXEL2 & 8 & 9 $\times$ 9 & 7 & 9 $\times$ 9 \\ \hline
PIXEL1 & 2.5 & 9 $\times$ 9 & 2.5 & 9 $\times$ 9 \\ \hline 
\end{tabular}
\end{center}
\caption{Hit position resolution of SSD + HFT layers for the two design versions. IST2 (simulated design) has two layers (A,B) with crossed strips.}
\label{tab:parameters}
\end{table}

\begin{figure}[htb]
\begin{minipage}[h]{0.49\textwidth}
\centering
\vspace{4pt}
\includegraphics[width=0.7\textwidth]{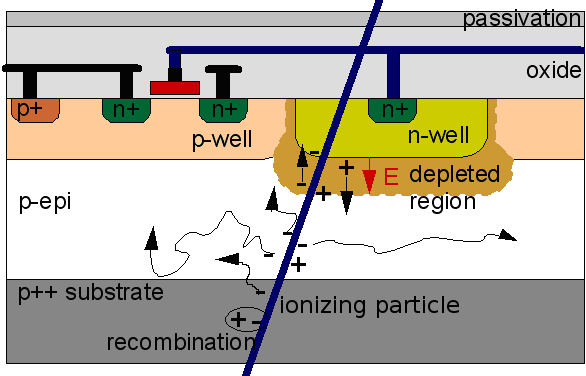}
\vspace{-3pt}
\caption{Electrons created in the epitaxial layer thermally diffuse towards low potential n-well region.}
\label{fig:maps}
\end{minipage}
\hfill
\begin{minipage}[h]{0.47\textwidth}
\centering
\vspace{4pt}
\includegraphics[width=0.93\textwidth]{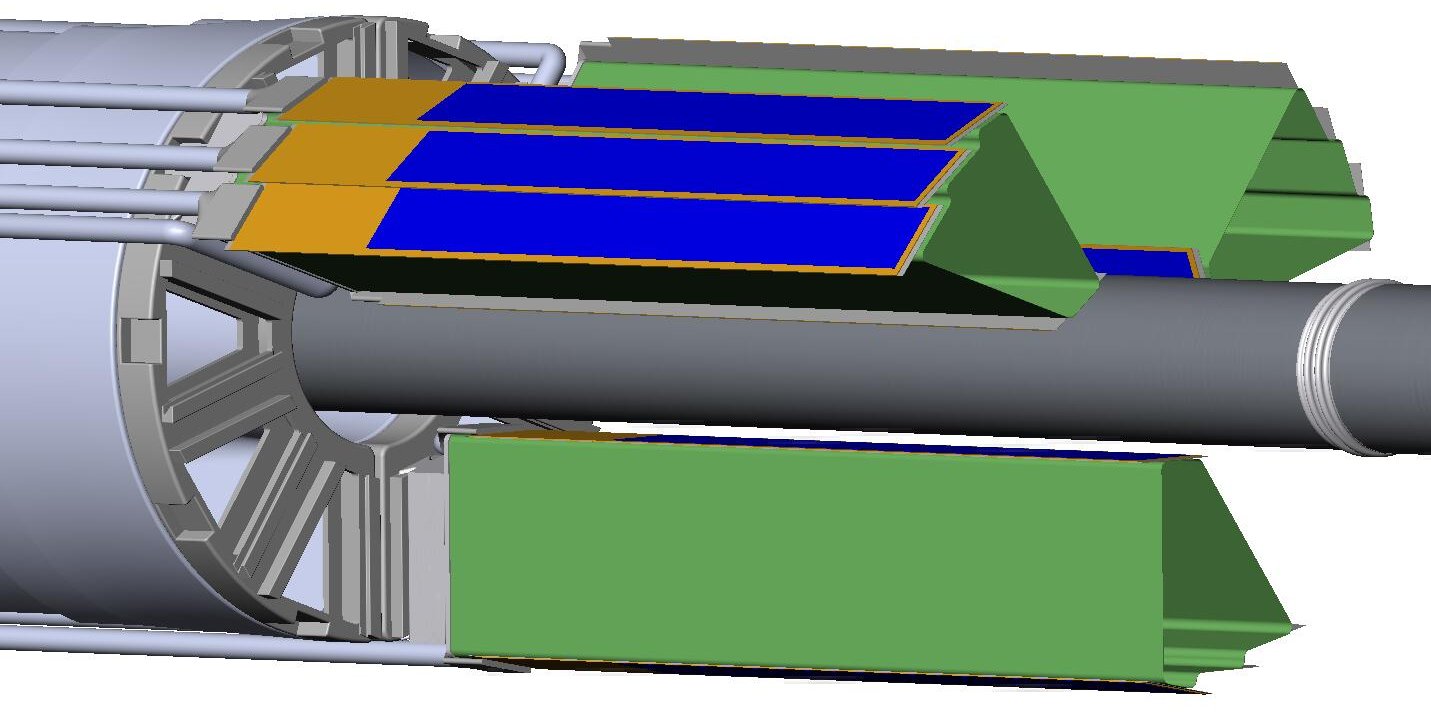}
\vspace{-3pt}
\caption{Detail of PIXEL detector patch design. 3 of the total 10 modules will be installed.}
\label{fig:patch}
\end{minipage} 
\hfill
\end{figure}

The PIXEL detector is made of low-mass monolithic active pixel sensors (MAPS, see Fig.~\ref{fig:maps}) and enables high precision measurement close to primary collision vertex, featuring 30 $\mu$m pixel pitch and thickness only $0.28 \%~X_{0}$ per layer. As there is a very large number of pixels, the readout of the PIXEL detector is relatively slow. 
Phase 1 sensors, which are planned to be used for a detector patch (Fig.~\ref{fig:patch}), will use a digital readout (LVDS) with an integration time of 640~$\mu$s. Integrating cluster finding into the sensor chip will allow for shorter ($<200~\mu\mathrm{s}$) integration times for the Ultimate sensors, to be used for the final detector.

Current small-size prototype sensors (MimoSTAR2) have been used to build a prototype telescope that was tested \cite{test} in the STAR experiment during the 2007 RHIC run (Au+Au, $\sqrt{s_{NN}} = 200~\mathrm{GeV}$). The test included integration with the STAR trigger system and confirmed a successfull operation of the system in the STAR experimental environment. 
   
Given the high luminosity projected for the future RHIC-II upgrade (up to $80 \cdot 10^{26} \mathrm{cm}^{-2} \mathrm{s}^{-1}$), the PIXEL detector will integrate over up to 10 minimum bias collisions. The IST detector is essential to improve hit identification at PIXEL2 layer in this pile-up environment. The IST detector consists of 1 layer of single-sided strip sensors, with very short ($<1~\mu\mathrm{s}$) integration time.

\section{Simulation results}

A full GEANT simulation of the STAR detector with HIJING central Au+Au events has been performed, assuming pile-up in the PIXEL detector at a level of RHIC-II luminosity. Pseudo-random hits were added to PIXEL1 and PIXEL2 layer, increasing hit density by a factor of 3. Single-particle efficiency (Fig.~\ref{fig:single_eff}) and pointing resolution (Fig.~{\ref{fig:pointing}) are generally in agreement with hand calculations.

\begin{figure}[htb]
\begin{minipage}[h]{0.48\textwidth}
\centering
\vspace{4pt}
\includegraphics[width=0.9\textwidth]{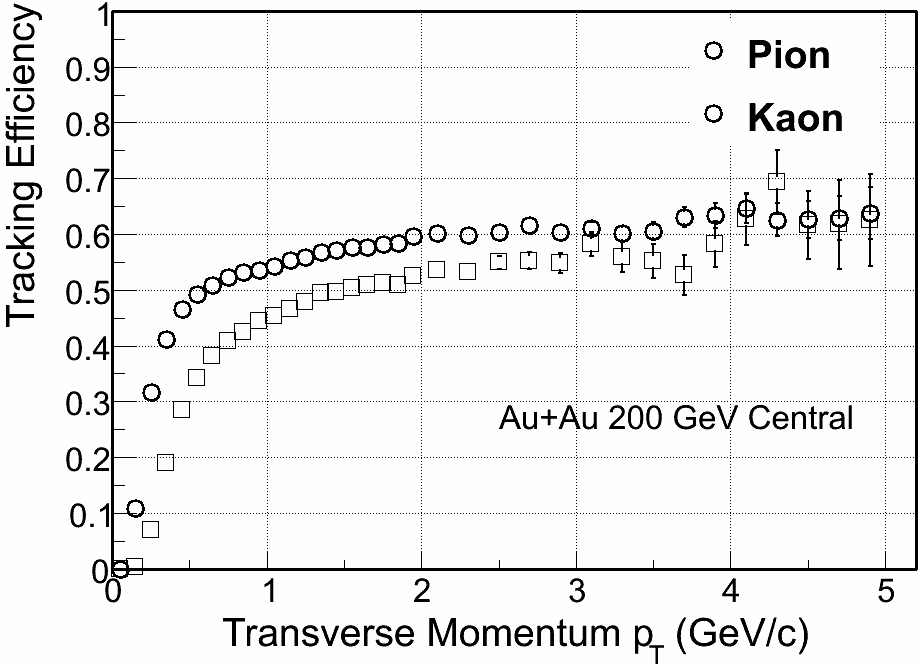}
\vspace{-4pt}
\caption{Reconstruction efficiency of pions and kaons in $|\eta|<1$ with correctly associated hits in both PIXEL layers.}
\label{fig:single_eff}
\end{minipage}
\hfill
\begin{minipage}[h]{0.48\textwidth}
\centering
\vspace{4pt}
\includegraphics[width=0.91\textwidth]{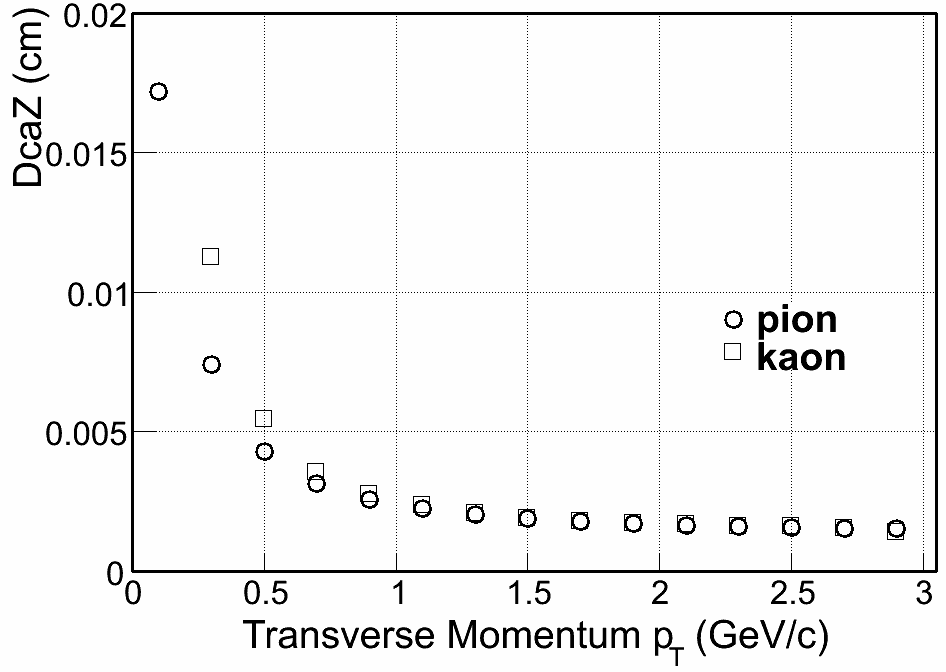}
\vspace{-4pt}
\caption{Pointing resolution in z-direction to primary vertex for tracks with hits in both PIXEL layers.}
\label{fig:pointing}
\end{minipage} 
\hfill
\end{figure}

To demonstrate the physics capabilities of the HFT detector, $\mathrm{D}^{0}$ mesons were embedded into HIJING events and reconstructed using $\mathrm{K}^{-}+\pi^{+}$ decay channel (B.R. 3.8\%). Reconstruction efficiency is shown in Fig.~\ref{fig:d0eff}. Small efficiency at low $\pt$ is caused by the short mean decay length of $\mathrm{D}^{0}$ ($123~\mu \mathrm{m}$). 
Signal significance of $\mathrm{D}^{0}+\overline{\mathrm{D}^{0}}$ from 100M central Au+Au collisions is shown in Fig.~\ref{fig:ss}. For realistic PID of daughter particles, kaon-pion separation up to $\pt \approx 1.6~\mathrm{GeV/}c$ is expected using the MRPC TOF detector (STAR upgrade to be finished in 2009). 
The conservative estimate of yield ($dN/dy = 0.002$ in p+p collisions) and power-law $\pt$ spectrum with $\langle \pt \rangle = 1.0~\mathrm{GeV/}c$ (consistent with p+p measurement and high $\pt$ suppression equal to that of charged hadrons \cite{jaro2}) was assumed.  

For measurements of $v_{2}$ and $R_\mathrm{CP}$, statistics of 500M minimum bias Au+Au events was assumed. This is the amount of data STAR detector can take in about 1 month of running (DAQ rate 500~Hz with DAQ1000 upgrade to be finished in 2008, and 40\% beam up time).
To estimate $\mathrm{D}^{0}$ and background yield at different centralities, $N_{bin}$ scaling for $\mathrm{D}^{0}$ and $(N_{part})^{2}$ scaling (pairs of particles) for background was used.
Estimated statistical errors (for $\mathrm{D}^{0}+\overline{\mathrm{D^{0}}}$, $|\eta|<1$) are shown in Fig.~\ref{fig:v2} ($v_{2}$) and Fig.~\ref{fig:Rcp} ($R_\mathrm{CP}$). The HFT will be able to distinguish between extreme elliptic flow scenarios for $\pt > 1.0~\mathrm{GeV/}c$ and measure $R_\mathrm{CP}$ with high precision for $\pt < 10~\mathrm{GeV/}c$ in the first year of operation.

\begin{figure}[htb]
\begin{minipage}[h]{0.48\textwidth}
\centering
\vspace{6pt} 
\includegraphics[width=0.9\textwidth]{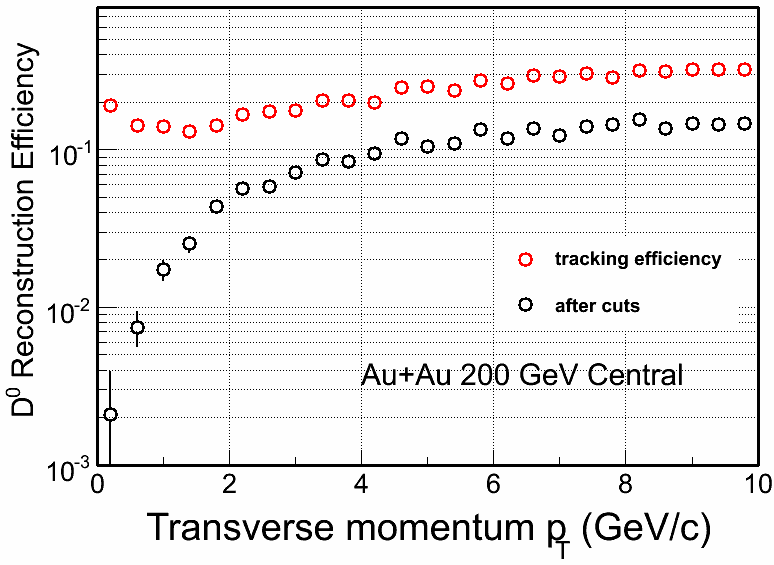}
\vspace{-7pt}
\caption{Efficiency to reconstruct $\mathrm{D}^{0}$ decay daughters and $\mathrm{D}^{0}$ (after applying cuts). For $\mathrm{D}^{0}$ in $|\eta|<1$.
}
\label{fig:d0eff}
\end{minipage}
\hfill
\begin{minipage}[h]{0.48\textwidth}
\centering
\vspace{4pt}
\includegraphics[width=0.9\textwidth]{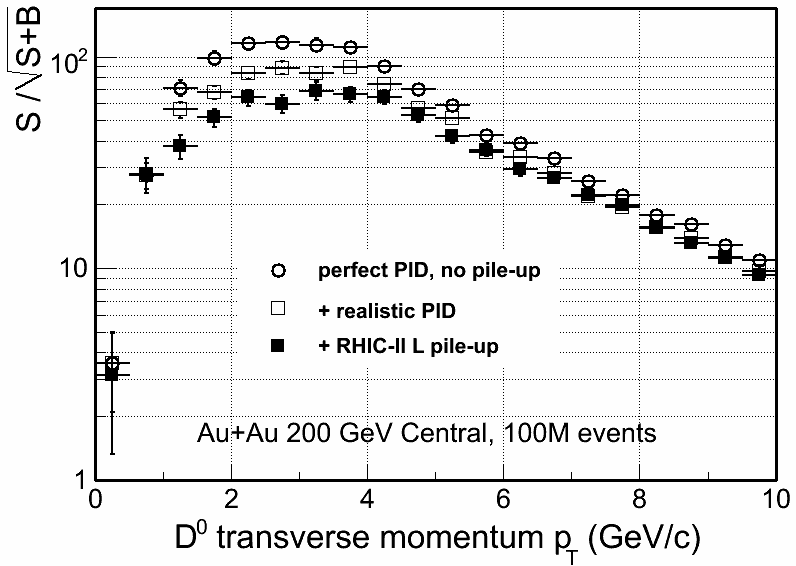}
\caption{$\mathrm{D}^{0}$ signal significance in central Au+Au events under different assumptions.
}
\label{fig:ss}
\end{minipage} 
\hfill
\end{figure}

\begin{figure}[htb]
\begin{minipage}[h]{0.48\textwidth}
\centering
\vspace{4pt}
\includegraphics[width=0.9\textwidth]{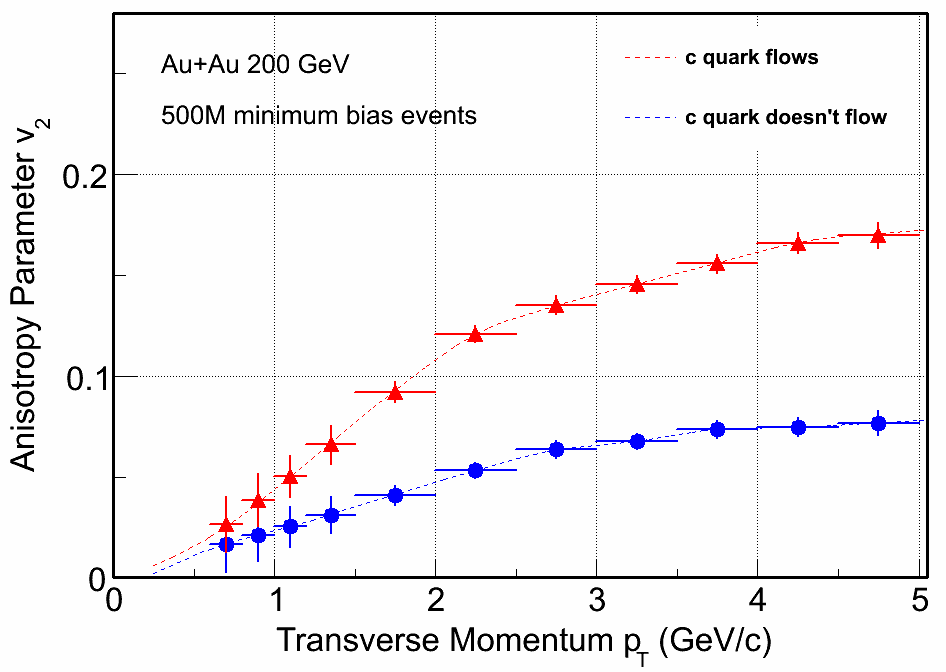}
\vspace{-4pt}
\caption{Two extreme scenarios for $\mathrm{D}^{0}$ elliptic flow and estimated statistical errors.}
\label{fig:v2}
\end{minipage}
\hfill
\begin{minipage}[h]{0.48\textwidth}
\centering
\vspace{4pt}
\includegraphics[width=0.9\textwidth]{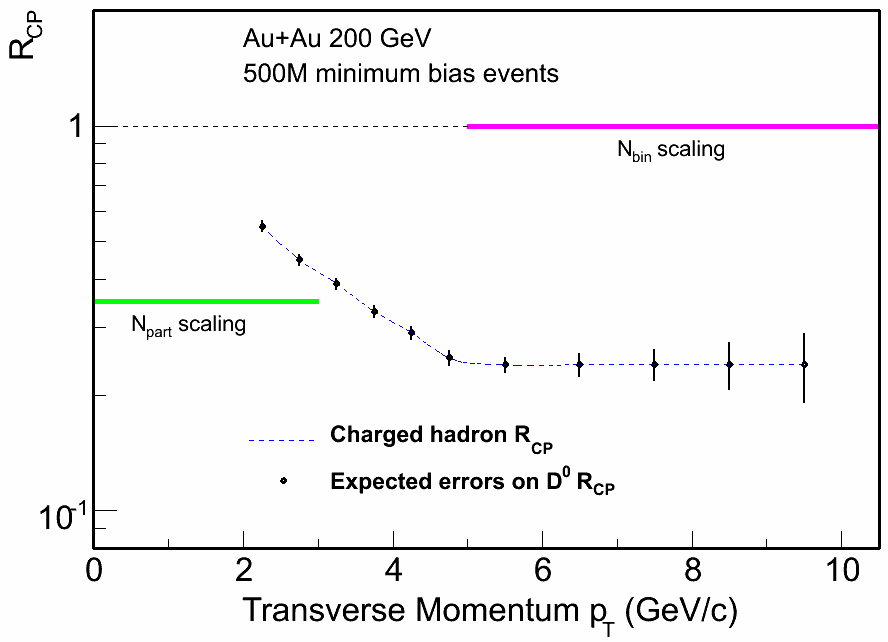}
\vspace{-4pt}
\caption{Estimated statistical errors of $R_\mathrm{CP}$ measurement for 1 month of data taking.}
\label{fig:Rcp}
\end{minipage} 
\hfill
\end{figure}

\section{Summary}

The HFT detector will measure open charm hadrons over a broad $\pt$ range, enabling precision study of charm quark collectivity and energy loss. These are important ingredients for a systematic study of the dense medium created in heavy-ion collisions at RHIC.

This will be achieved by using low mass MAPS sensors (PIXEL) together with a fast strip detector (IST), delivering ultimate pointing resolution at low $\pt$, even in the high luminosity environment of RHIC-II.

MAPS prototype telescope was successfully tested during 2007 Au+Au run, making important step in the PIXEL detector development.

\section*{Acknowledgements}
This work was supported in part by the IRP AV0Z10480505, 
by GACR grant 202/07/0079 and by grant LC07048 of the Ministry 
of Education of the Czech Republic.

\end{document}